\documentclass[12pt]{article}
\usepackage{amssymb}
\usepackage{epsfig}
\usepackage{graphicx}
\usepackage{amsmath}
\usepackage{verbatim}

\csname @addtoreset\endcsname{equation}{section}
\begin{document}
\begin{titlepage}
\title{\bf\Large A Simple Model of Direct Gauge Mediation \vspace{18pt}}

\author{\normalsize Sibo~Zheng and Yao~Yu\vspace{12pt}\\
{\it\small   Department of Physics, Chongqing University, Chongqing 401331, P.R. China}\\}

\date{}
\maketitle \voffset -.3in \vskip 1.cm \centerline{\bf Abstract}
\vskip .3cm  In the context of direct gauge mediation Wess-Zumino models  are very attractive in supersymmetry model building.
Besides the spontaneous supersymmetry and $R$-symmetry breaking,
the problems of small gaugino mass as well as $\mu$ and B$\mu$ terms
should be solved so as to achieve a viable model.
In this letter, we propose a simple model as an existence proof, 
in which all these subjects are realized simultaneously,  with no need of fine tuning. This completion also implies that much of parameter space for direct gauge medition can be directly explored at  LHC.
\vskip 5.cm \noindent 08/ 2011
 \thispagestyle{empty}

\end{titlepage}
Supersymmetry  (SUSY) is an appealing candidate for explaining the mass hierarchy  and providing the unification of gauge couplings. 
Experimental searches at colliders such as LHC give arise to strigent constraints on supersymmetric physical parameters at low energy where SUSY must be broken.  
Some difficulties can be avoided by adjusting the  mechanism of SUSY breaking or allowing a few fine tunings. 

On the realm of SUSY model building,  gauge mediated SUSY breaking , is one of the most well studied scenarios for a few reasons.
At first, the problem of flavor changing neutral currents can be naturally solved with supersymmetric particles $\sim 1$TeV in this paradigm. 
Second, this makes the supersymmetry testable at a few current colliders such as LHC.
Finally, microscopic models which trigger SUSY breaking can be constructed in a wide SUSY theories.

In particular, chiral theories, or concretely Wess-Zumino models \cite{0710.3585} are very attractive and powerful.
One reason is that they often serve as the effective theories of (strongly coupled ) microscopic theories,
for example the well known massive SQCD theory with a dual description \cite{ISS}.
Furthermore, they are calculable and under control, thus might be applied to viable SUSY phenomenology.
Very recently, it is generally argued that Wess-Zumino models
in which SUSY is spontaneously broken are actually type of O' Raifeartaigh models \cite{Shih09},
\begin{eqnarray}{\label{e1}}
W=fX+(\lambda_{ij}X+m_{ij})\varphi_{i}\tilde{\varphi}_{j}+\cdots
\end{eqnarray}
where neglected terms denote the cubic terms.

What is of more interest is to directly apply these O' Raifeartaigh models
to models buildings in the context of gauge mediation, i.e, direct gauge mediation (DGM)
\cite{DGM1.1, DGM1.2,DGM1.3}.
In contrast with the minimal gauge mediation (see review \cite{Giudice} and references therein),
the is no need to introduce additional messenger sector in DGM.
At first sight, it is observed that the gaugino mass of order $\mathcal{O}(F)$
( $\sqrt{F}$ refers to the supersymmetry breaking scale)
often vanishes in direct gauge mediated O' Raifeartaigh models,
Now it is understood \cite{Shih09} that this phenomena
is tied to the global vacuum structure composed of pseudomoduli space $X$,
whether $R$-symmetry is spontaneous broken or not.
In light of this new finding,
various O' Raifeartaigh models where the gaugino mass problem can be resolved are proposed
\cite{DGM2,DGM3,DGM4,DGM5,DGM6,DGM7,DGM8, DGM9, DGM10},
some of which even have microscopic completions \cite{DGM4, DGM10}.

In this letter, we discuss another important subject left in DGM,
that is the generation of viable $\mu$ and B$\mu$ terms as in ordinary gauge mediation
\cite{mu1,mu2,mu3,mu4,mu5,Csaki,1101.3827}
\footnote{In \cite{Csaki}, the authors discuss the strongly coupled generation of $\mu$ term in context of direct
gauge mediation.}.
This subject have been less addressed in comparison with the gaugino mass problem.
We restrict us to most generalized O' Raifeartaigh models that respect renormalization and $R$-symmetry,
and then address the gaugino mass and $\mu$ problem simultaneously.

First, what kind of O' Raifeartaigh models in \eqref{e1} can solve the gaugino mass problem ?
We can take a few limits in \eqref{e1} for illustration.
If $\lambda$ is diagonal (via bi-unitary transformation ) and $m=0$,
this actually reduces to the minimal gauge mediation, in which det$\mathcal{M}=X^{N}$det$\lambda$.
As well known there is no small gaugino mass problem in this context .
However, it is nerve considered as starting point of direct gauge mediation,
as spontaneous SUSY breaking can not be realized in this setup.
If $m$ is diagonal then det$\mathcal{M}=$det$m$
\footnote{The field space composed of the pseudomoduli $X$ is stable globally in this type of O' Raifeartaigh models,
which implies that the determinant $\mathcal{M}$ is a constant \cite{Shih09}.},
which results in spontaneously broken SUSY and vanishing gaugino mass at order of $\mathcal{O}(F)$.

Therefore, in order to render the O' Raifeartaigh model to generate the one-loop gaugino masses,
or equivalently guarantee the determinant to depend on $X$,
there must be at least one non-zero diagonal element in $\lambda$.
So the superpotential can be constructed as
the mixing of those of minimal setup and tree-level mass terms of messengers.

Now, we consider a concrete model in light of above observations,
whose superpotential is given by,
\begin{eqnarray}{\label{e2}}
W_{1}&=&fX+\lambda_{1}X\left(S\tilde{S}+T\tilde{T}\right)+m_{1}S\tilde{T}\nonumber\\
&+&\lambda_{2}X\left(\varphi_{1}\tilde{\varphi}_{1}+\varphi_{3}\tilde{\varphi}_{3}\right)+m_{2}\varphi_{1}\tilde{\varphi}_{3}
+\lambda_{3}X\left(\varphi_{2}\tilde{\varphi}_{2}+\varphi_{4}\tilde{\varphi}_{4}\right)+m_{3}\varphi_{2}\tilde{\varphi}_{4}
\end{eqnarray}
This is the minimal setup as we will find.
We assume all the masses and couplings in \eqref{e2} are real without loss of generality.
The couplings in \eqref{e2} can be realized via imposing global symmetries $[SU(2)\times~SU(2)]^{2}$ as follows,
\begin{eqnarray}{\label{e3}}
\Phi=\left(\begin{array}{c}\varphi_{1} \\ \varphi_{2}\end{array}\right),~~~~
\tilde{\Phi}=\left(\begin{array}{c}\tilde{\varphi}_{1} \\ \tilde{\varphi}_{2}\end{array}\right),~~~~
\Sigma=\left(\begin{array}{c}\varphi_{3} \\ \varphi_{4}\end{array}\right),~~~~
\tilde{\Sigma}=\left(\begin{array}{c}\tilde{\varphi}_{3} \\ \tilde{\varphi}_{4}\end{array}\right).
\end{eqnarray}
Also the global symmetry assignment results in the degeneracies $\lambda_{2}=\lambda_{3}$ and $m_{2}=m_{3}$.
Thus, eq\eqref{e2} can be rewritten  as,
\begin{eqnarray}
W_{1}&=&fX+\lambda_{1}X\left(S\tilde{S}+T\tilde{T}\right)+m_{1}S\tilde{T}
\nonumber\\
&+&\lambda_{2}X\left(\Phi\tilde{\Phi}+\Sigma\tilde{\Sigma}\right)
+m_{2}\Phi\tilde{\Sigma}
\end{eqnarray}

We assume $S,\tilde{S}$ and $T,\tilde{T}$ as standard model singlets.
Additional coupling associated with the Higgs fields can be introduced
when the $SU(2)$ global symmetries are directly gauged as the standard model electroweak groups,
\begin{eqnarray}{\label{e4}}
W_{2}=\lambda_{\mu}\tilde{S}\tilde{\Phi} H_{\mu}+\lambda_{d}S\Phi H_{d}
\end{eqnarray}
In particular, either $S,\tilde{S}$ fields or $T,\tilde{T}$ can couple to the Higgs doublets,
but they can not be allowed to appear in \eqref{e4} at same time as a result of $R$-symmetry. \footnote{
The reason is due to the absence of quardratic mass terms for $S^2$ and  $T^{2}$ in \eqref{e2}. We undersatand this fact as a consequence of  $R(S)\neq 1$ and $R(T)\neq 1$. Similar understanding can also be applied to  singelt fields $\tilde{S}$ and $\tilde{T}$.}
Similarly, either $\Phi, \tilde{\Phi}$ or $\Sigma,\tilde{\Sigma}$ can be coupled to Higgs fields.
We choose the set in \eqref{e4} for example.
The superpotential of O' Raifeartaigh models we consider is $W=W_{1}+W_{2}$,
which respects gauge symmetries of standard model and $R$-symmetry involved.
Once all the subjects involved in SUSY model buildings are realized in such kind of models,
one can add triplet fields of QCD gauge group in \eqref{e2} so as to complete the model.

According to \eqref{e2} and \eqref{e4},
the vacuum is represented by,
\begin{eqnarray}{\label{e5}}
S=\tilde{S}=0,
~~~~\left(\begin{array}{c}
                    \Phi \\
                    \Sigma
                  \end{array}\right)=0,~~~~~~
                  \left(\begin{array}{c}
                    \tilde{\Phi} \\
                    \tilde{\Sigma}
                  \end{array}\right)=0
,~~~ X~~arbitrary
\end{eqnarray}
with potential $V=f^{2}$.
To achieve this vacuum,
the property that there are diagonal $\lambda$ and non-zero mass terms in \eqref{e2} is crucial in above analysis.
At this SUSY breaking vacuum \eqref{e5} the gauge symmetries of standard model is unbroken.
Even without studying the details of pseudomoduli space $X$,
one expects that there is no gaugino mass problem in this model.
Since in the region $X\rightarrow 0$, some freedoms in messengers become tachyonic.
This means that the vacuum \eqref{e4} is not stable globally.
From \eqref{e2}, the eigenvalues of messenger fermion mass squared $\mathcal{M}^{2}_{F}$ are given by,
\begin{eqnarray}{\label{e7}}
m^{2}_{1/2,i^{\pm}}=m^{2}\left(\frac{1}{2}+x_{i}^{2}\pm\sqrt{\frac{1}{4}+x_{i}^{2}}\right)
\end{eqnarray}
for a given basis $i$. In \eqref{e7} we have defined the dimensionless coefficients $x_i=\lambda_{i}X/m_{i}$.
Similarly, it is straightforward to evaluate the messenger boson mass squared $\mathcal{M}^{2}_{B}$.
From these eigenvalues we verify that some fermions are massless while some bosons tachyonic at small $X<\sqrt{f}$.
So the physical parameter space is given by
\begin{eqnarray}{\label{n1}}
\sqrt{f}<< X < \min{(m_{i})}
\end{eqnarray}
Actually, a ratio of order $\mathcal{O}(10)$ for the first constraint in \eqref{n1}
is sufficient to guarantee the positive masses of messenger fields.
In this note,  we will take the small $F$ limit
in order to simplify the analysis of Coleman-Weinberg potential in the next paragraph.

Let us examine the $R$-symmetry breaking in our model.
In \eqref{e2} one finds that there must be $R$-charge assignments other than 0 or 2 in \eqref{e2}.
Following the argument in \cite{Shih07},
which states that $R$-symmetry can not be broken except there are fields with $R$-charge other than 0 and 2,
one can see that the
$R$-symmetry breaking or equivalently negative mass squared $m_{X}^{2}$ is not difficult to be realized.
According to discussions in the previous paragraph,
it is sufficient to study the region of moderate $X$ value,
we will focus on this region with small $F$-term.
Under limit \eqref{n1} the one-loop Coleman-Weinberg potential $V_{CW}(X)$ for the pseudomoduli
at moderate $X$ is approximately given by,
\begin{eqnarray}{\label{e8}}
V_{CW}(X)=\frac{5f^{2}}{32\pi^{2}}\sum_{i=1}^{i=3}\lambda_{i}^{2}V_{2}(x_{i}),~~~~x_i=\lambda_{i}X/m_{i}
\end{eqnarray}
where
\begin{eqnarray}{\label{e9}}
V_{2}(x_{i})=-\frac{2}{1+4x_{i}^{2}}+4\log~x_{i}+\frac{2x_{i}^{2}+1}
{\left(4x_{i}^{2}+1\right)^{\frac{3}{2}}}\log\frac{2x^{2}_{i}+1+\sqrt{4x_{i}^{2}+1}}{2x^{2}_{i}+1-\sqrt{4x_{i}^{2}+1}}
\end{eqnarray}
The Coleman-Weinberg potential is plotted in fig. 1;
one finds that $V_{CW}$ is minimized at $x_{1}=x_{2}=x_{3}\simeq 0.25$  or $X_{0}\sim 0.1$ m.
\begin{figure}[h!]
\centering
\begin{minipage}[b]{0.8\textwidth}
\centering
\includegraphics[width=3.5in]{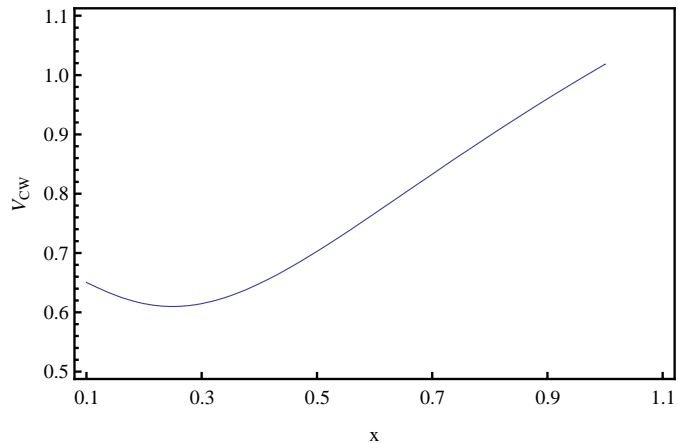}
\end{minipage}%
\caption{$V_{CW}$ varies as function of $x_{i}$ in unit of $f^{2}$.
For illustration,
take the particular values $\lambda_{1}=\lambda_{2}=\lambda_{3}=3$ and $m_{1}=m_{2}=m_{3}=$ m. }
\end{figure}

\begin{figure}[h!]
\centering
\begin{minipage}[b]{0.45\textwidth}
\centering
\includegraphics[width=2.7in]{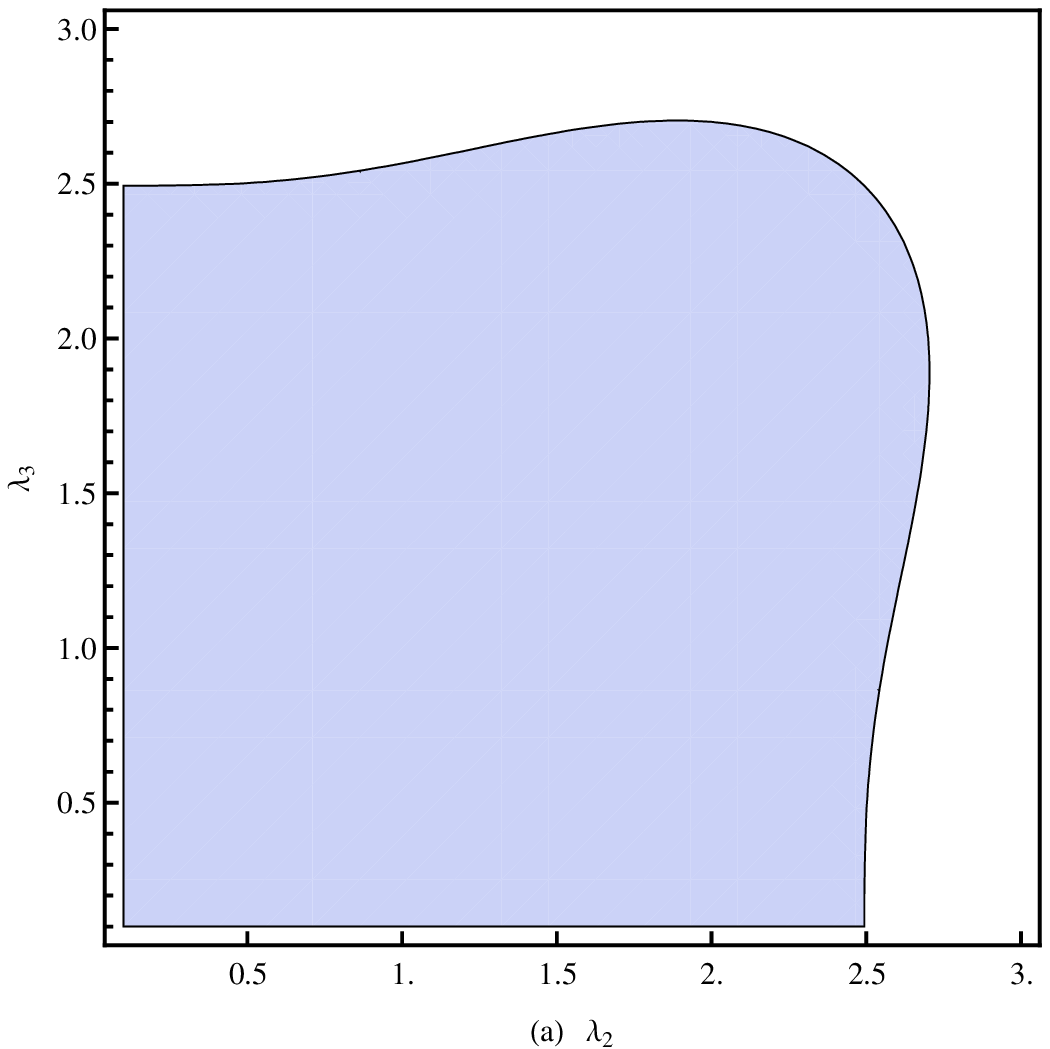}
\end{minipage}%
\begin{minipage}[b]{0.45\textwidth}
\centering
\includegraphics[width=2.7in]{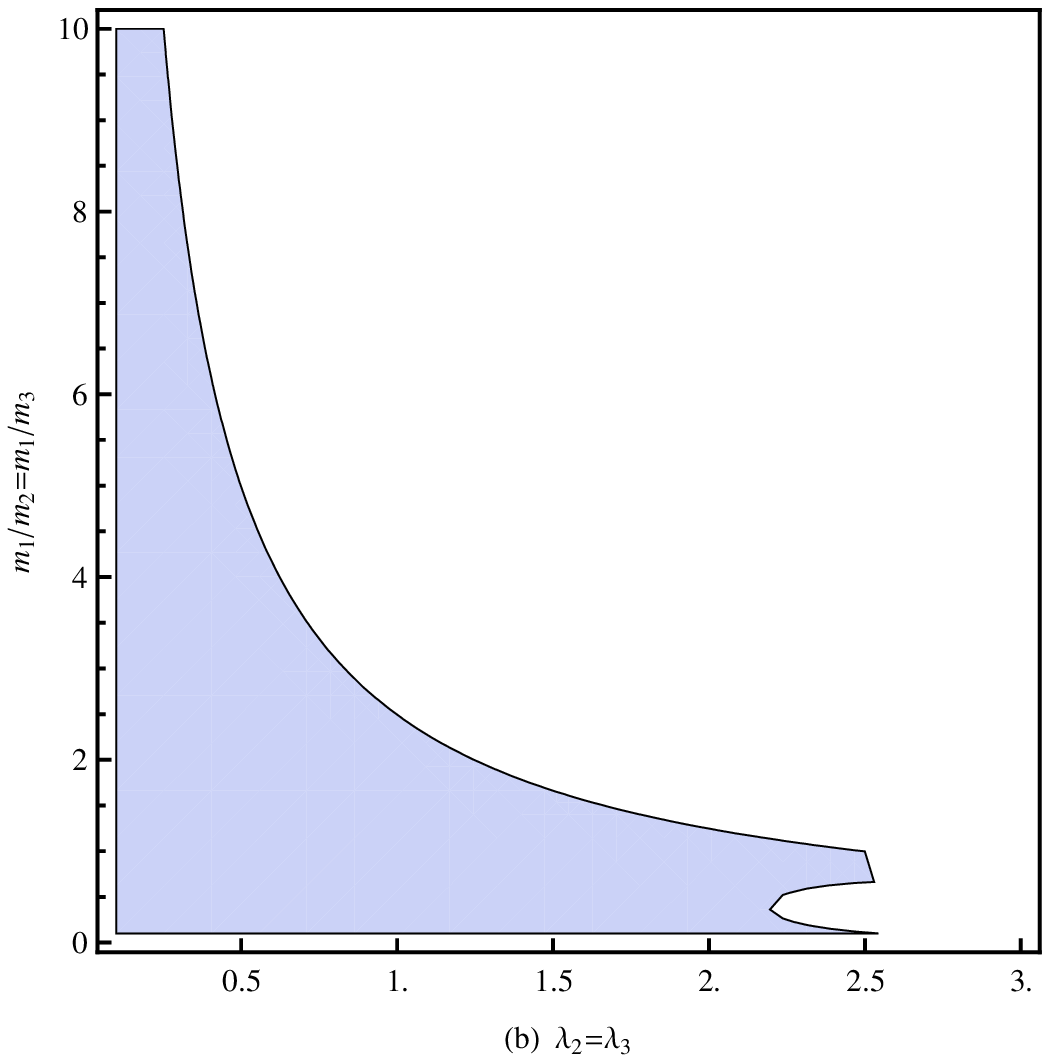}
\end{minipage}
\caption{Unified masses $m_{i}=$m and $X_{0}=0.1$ m in $(a)$.
The parameter space composed of $\lambda_{2}$ and $\lambda_{3}$ is shown in the region $0.1\leq\lambda_{1}\leq 3$.
Non-degenerate masses among $m_{2}$ and $m_{1}$ in $(b)$.
We set the messenger scale $X_{0}=0.1 m_{1}$, while $\lambda_{1}$ varies in the region $0.1-3$. }
\end{figure}

Firstly, we set all masses $m_i$ are unified in order to simplify the analysis.
As shown in fig. 2$(a)$, if one wants to obtain $X_{0}= 0.1 m_{1}$,
$\lambda_{2}$ and $\lambda_{3}$ should be chosen around the window $0.1-2.5$,
when $\lambda_{1}$ varies from 0.1 to 3
\footnote{Large Yukawa couplings often give rise to the problem of Landau pole in the context of
direct gauge mediation \cite{Poppitz, 0812.4600}}.
If $X_{0}\leq 0.01$ m, we find that there are no parameter space allowed.
Relax the condition $m_{i}=$ m and allow deviation of $m_{2}$ from $m_{1}$,
we show in fig. 2.$(b)$ the parameter space when $\lambda_{2}=\lambda_{3}$, $X_{0}= 0.1 m_{1}$
and $\lambda_{1}$ varies from 0.1 to 3.

Now we proceed to discuss the soft terms induced by superpotential \eqref{e4},
which can be read from the one-loop effective Kahler potential $K_{eff}$
after integrating out the messenger fields involved \cite{Kahler},
\begin{eqnarray}{\label{e10}}
K_{eff}=-\frac{1}{32\pi^{2}}Tr\left(\mathcal{M}^{\dag}\mathcal{M}
\log\frac{\mathcal{M}^{\dag}\mathcal{M}}{\Lambda^{2}}\right)
\end{eqnarray}
From \eqref{e4}, in the case $m_{3}=m_{2}$ and $\lambda_{3}=\lambda_{2}$ matrix
$\mathcal{M}^{\dag}\mathcal{M}$ is reduced to $4\times4$ and given by,
\begin{eqnarray}{\label{e11}}
\mathcal{M}^{\dag}\mathcal{M}=\left(
\begin{array}{cccc}
 \lambda_{2}^{2}\mid X\mid^{2}+\lambda_{\mu}^{2}\mid H_{\mu}\mid^{2} & \lambda_{2}\lambda_{\mu}X^{*}H_{\mu}+\lambda_{1}\lambda_{\mu}XH_{\mu}^{*} & 0 & \lambda_{2}m_{2}X^{*}\\
\lambda_{2}\lambda_{d}XH^{*}_{d}+\lambda_{1}\lambda_{\mu}X^{*}H_{\mu} & \lambda_{1}^{2}\mid X\mid^{2}+\lambda_{d}^{2}\mid H_{d}\mid^{2}+m_{1}^{2}&\lambda_{1}m_{1}X & \lambda_{d}m_{2}H_{d}^{*}\\
0&\lambda_{1}m_{1}X^{*}&\lambda_{1}^{2}\mid X\mid^{2}&0\\
\lambda_{2}m_{2}X & \lambda_{d}m_{2}H_{d} &0 &\lambda^{2}_{2}\mid X\mid^{2}
\end{array}
\right)
\end{eqnarray}
under basis $\left(\Phi, \tilde{S}, \tilde{T}, \Sigma\right)\mathcal{M}
\left(\tilde{\Phi}, S, T, \tilde{\Sigma}\right)^{T}$.

Relevant mass terms can be read  from \eqref{e10} as,
\begin{eqnarray}{\label{A}}
\mu&=&\frac{\partial}{\partial\bar{\theta}^{2}}\mathcal{Z}_{\mu d}\mid_{\theta=\bar{\theta}=0},\nonumber\\
B\mu&=&-\frac{\partial}{\partial\bar{\theta}^{2}}\frac{\partial}{\partial\theta^{2}}\mathcal{Z}_{\mu d}\mid_{\theta=\bar{\theta}=0}\nonumber\\
m^{2}_{H_{\mu}}&=&-\frac{\partial}{\partial\bar{\theta}^{2}}\frac{\partial}{\partial\theta^{2}}\log \mathcal{Z}_{\mu }\mid_{\theta=\bar{\theta}=0}\\
m^{2}_{H_{d}}&=&-\frac{\partial}{\partial\bar{\theta}^{2}}\frac{\partial}{\partial\theta^{2}}\log \mathcal{Z}_{d }\mid_{\theta=\bar{\theta}=0}\nonumber
\end{eqnarray}
where $Z_{\mu d}$ and $Z_{\mu,d}$ are given by,
\begin{eqnarray}{\label{A0}}
Z_{\mu d}&=&\frac{\partial}{\partial( H_{\mu}H_{d})}K_{eff}\mid_{H_{\mu}=H_{d}=0},\nonumber\\
Z_{\mu, d}&=&\frac{\partial}{\partial( H^{\dag}_{\mu,d}H_{\mu,d})}K_{eff}\mid_{H_{\mu,d}=H^{\dag}_{\mu, d}=0}
\end{eqnarray}
Since these soft terms are generated through one hidden sector in our framework,   
our  model belongs to what is known as one-scale gauge medaition.  
As discussed in \cite{0812.3900},  one roughly expects a relation as
\begin{eqnarray}
\mid B\mu \mid \sim m^{2}_{H_{\mu,d}} >> \mu^{2}
\end{eqnarray}
which plagues these one-scale models and indicates the failure of EWSB.
However, more precise estimates needs to be done so as  to verify this relation given a specific model,
and it is not impossible to avoid this relation in some circumstances.

Here we point out some possibilities.
One choice is that  $m^{2}_{H_{\mu}}$ is  negative, 
with its absolute value  smaller than positive $m^{2}_{H_{d}}$ but larger than $\mu^{2}$. 
Another choice is that one allows a large $m^{2}_{H_{d}}$ and small $m^{2}_{H_{\mu}}$, 
with a small hierarchy  $m^{2}_{H_{d}}>> B\mu $ so that it can balance the influence coming from the small hierachy $B\mu>>\mu^{2}$  \cite{0809.4492}. 
We refer \cite{1106.5553} to the reads  for more discussions about this issue. 
As we will see the model we discuss here is a new example in the first choice.

Since the matrix \eqref{e11} is quite complicated
so that the effective Kahler potential can not be generally evaluated,
we take the limit $m_{2}=m_{3}$ and $\lambda_{2}=\lambda_{3}$ to simplify the simulation.
Note that these choices correspond to a favored parameter space, as seen in fig 2.$(b)$.

The leading contributions to $m^{2}_{H_{\mu,d}}$ are composed of two parts.
One arises from the ordinary gauge mediation.
The other comes from the superpotential \eqref{e4}.
The later contribution induced at one-loop,   generally dominates over the former. 
By using the conditions of electroweak symmetry breaking,
\begin{eqnarray}{\label{e15}}
(c.1)&:&~~ (B\mu)^{2}>(\mid\mu\mid^{2}+m^{2}_{H_{\mu}})(\mid\mu\mid^{2}+ m^{2}_{H_{d}})\nonumber\\
(c.2)&:&~~2B\mu<2\mid\mu\mid^{2}+m_{H_{\mu}}^{2}+m_{H_{d}}^{2}
\end{eqnarray}
For $\lambda_{2}=\lambda_{3}=1$  and fixed scale $X=0.1m_{2}$ ,
 it turns out that the allowed parameter space is given by 
\footnote{Since couplings $\lambda_{\mu}$ and $\lambda_{d}$ are overall coefficients in $\mu$ and $B\mu$ terms,
 we have taken $\lambda_{\mu}=\lambda_{d}=1$ for simplicity. 
Also note that large deviation from $\lambda_{2}\sim \lambda_{3}\sim 1$ is not consistent with the choice $X\sim 0.1$m, as shown in fig. 2. },
\begin{eqnarray}{\label{A2}}
\lambda_{1}\sim 0.26, ~~~~~  m_{1}/m_{2}\sim 0.12
\end{eqnarray}
which results in the following spectra in our model, 
\begin{eqnarray}{\label{A3}}
 m^{2}_{H_{\mu}}: \mu^{2}: B\mu : m^{2}_{H_{d}}\sim 1: 2: 500: 10^{3}
\end{eqnarray}
after we put values of \eqref{A2} into \eqref{e11} and \eqref{A}. 
The spectra \eqref{A3}  suggests that our model is an example of large $m_{H_{d}}$ and small $m_{H_{\mu}}$ mentioned above.

What about the RG effects on the spectra given by \eqref{e11}
when one runs from $X_{0}$ to the electroweak scale ?
Since there are no multiple messenger threshold corrections in our model,
the RG effects are quite simple.
According to the RG equations of MSSM given in \cite{gre},
one observes that the $m_{H_{\mu}}$ receive its quantum corrections more substantially
than $\mu$, $m_{H_{d}}$ and $B_{\mu}$. 
If we restrict us to low-scale gauge mediation with $X_{0}\sim 10^{3}-10^{7}$TeV, 
the correction can be estimated through linear approximation.
For the spectra given by \eqref{A3} ,
 $\delta m^{2}_{H_{\mu}}\sim -0.1\times m^{2}_{H_{d}}/16\pi^{2}\sim - \mu^{2}$.
This negative contribution implies that the first condition in \eqref{e15} can be still satisifed,
while the second condition does not substantially modified .
We refer the readers to the recent work \cite{1106.5553} on this subject through effective field theory analysis.

What about the other choices such as $m_{2}=m_{3}<< m_{1}$ and $\lambda_{2}=\lambda_{3}<< \lambda_{1}$,
or $m_{2}=m_{3}<< m_{1}$ and $\lambda_{2}=\lambda_{3}>> \lambda_{1}$, or
$m_{2}=m_{3}>> m_{1}$ and $\lambda_{2}=\lambda_{3}<< \lambda_{1}$ ?
We find that it is often impossible to both satisfy the electroweak symmetry breaking conditions $(c.1)$ and $(c.2)$ in
these cases.
What is worse is that the parameter space to generate the one-loop gaugino masses is substantially
suppressed under these limits, as shown in fig. 2$(b)$.\\

In summary, we propose a simple Wess-Zumino model,
which can serve as viable SUSY model of direct gauge mediation.
In this scenario, all messengers involve in supersymmetry breaking.
The $R$-symmetry is also spontaneously broken as a result of the specific choices of $R$-charges.
Phenomenologically, We find the gaugino mass is induced at one-loop, with the same order of the scalar masses.
Also, there is no $\mu$ problem associated with soft masses in Higgs sector,
which can be naturally solved in our model, 
with no need of fine tunings among Yukawa couplings in the SUSY breaking hidden sector.
Since most of supersymmetric particles $\sim 1$TeV, 
this makes part of direct gauge mediated SUSY models testable at LHC.

As far as we know in the literature,
our model is the first example that address all these important issues.
We do not discuss the possible searches of relevant signals at the LHC.
The main goal in this paper is to provide an  existence proof in SUSY model building in the context of direct gauge mediation.
\\

~~~~~~~~~~~~~~~~~~~~~~~~~~~~~~~~~~~~~~~~
$\bf{Acknowledgement}$\\
We thank Jia-Hui Huang for discussions and  Jin  Min Yang  for reading the manuscript.
This work is supported in part by the Fundamental Research Funds for the Central Universities with project
number CDJRC10300002.

\end{document}